\documentclass[graphicx,twocolumn,superscriptaddress]{revtex4}
\usepackage[pdftex]{graphicx}
\usepackage{xcolor}

\begin{document}


\title{Localization and magnetism of the resonant impurity states in Ti doped PbTe} 



\author{Bartlomiej Wiendlocha}
\email[e-mail: ]{wiendlocha@fis.agh.edu.pl}
\affiliation{AGH University of Science and Technology, Faculty of Physics and Applied Computer Science, Al. Mickiewicza 30, 30-059 Krakow, Poland}
\affiliation{Department of Mechanical and Aerospace Engineering, the Ohio State University, Columbus, Ohio 43210, USA}

\date{\today}

\begin{abstract}
The problem of localization of the resonant impurity states is discussed for an illustrative example of Titanium doped Lead Telluride. 
Electronic structure of PbTe:Ti is studied using first principles methods, densities of states and Bloch spectral functions are analyzed. We show that Ti creates resonant states in the conduction band of PbTe, however, spectral functions of the system strongly suggest  localization of these states and show poor hybridization with PbTe electronic structure. 
The contrast between results presented here and previously reported spectral functions for PbTe:Tl correlate very well with the different effect of those impurities on thermopower ($S$) of PbTe, which is large increase is $S$ for PbTe:Tl and almost no effect on $S$ for PbTe:Ti.
Moreover, magnetic properties of the system are studied and formation of magnetic moments on Ti atoms is found, both for ordered (ferromagnetic) and disordered (paramagnetic-like) phases,  {showing that PbTe:Ti can be a magnetic semiconductor.}
\end{abstract}


\maketitle 

{\it Introduction}
Localization or delocalization of the electronic states, introduced to the semiconductor upon doping with impurity atoms, is the key parameter for understanding the basic physical properties of the doped material. It is especially important for improving the efficiency of thermoelectric (TE) materials via doping with the resonant impurities,\cite{ees-review} since localized electronic states generally do not take part in transport phenomena, and resonant states may be predisposed for localization, as they introduce sharp peaks in the densities of electronic states.

In this work, by using Korringa-Kohn-Rostoker method with Coherent Potential Approximation (KKR-CPA), we address these topics by studying electronic structure of Titanium doped Lead Telluride, PbTe:Ti, and contrasting it to the recently published results for Thallium doped PbTe.\cite{bw2013} In both cases, impurity atoms, Ti and Tl, create resonant states in PbTe.\cite{ravich-review,ravich-review2} However, in the Tl doped PbTe, thermopower (and thermoelectric efficiency) was found to increase, comparing to other, non-resonant impurities.\cite{heremans-science,ees-review,ees-jaworski,pbte-tl-si-nano} On the other hand, in PbTe:Ti no increase in thermopower above the Pisarenko line (i.e. thermopower as a function of carrier concentration) for the n-type PbTe was found.\cite{konig-pbte_ti} This difference was tentatively explained before as the result of more localized nature of the resonant states for the PbTe:Ti case, since here resonant levels come from 3d electronic states of Titanium, more 
likely to be localized than 6s states of Thallium.\cite{ees-review,konig-pbte_ti}
In present work, we give strong support to the localization of Titanium impurity states and show that the KKR-CPA method allows to study such effects. Calculated Bloch spectral functions, which describe the dispersion relations for the disordered system, show that Ti 3d states are likely to form dispersionless, flat and narrow impurity band, characteristic for a localized states. Also, the electronic states which appear in the Brillouin zone of PbTe upon substitution of Pb with Ti show very small hybridization with Te electronic states, confirming the isolated character of the impurity.
The formation of resonant states on transition-metal impurities in PbTe was reported in literature (e.g. Cr\cite{pbte_cr_jap,pbte_cr_apl,Grodzicka1994,michele-pbtecr,story-pbtecr}, Fe\cite{pbte_fe-heremans,pbte_fe_press}, Sc\cite{pbte_sc}) however, the problem of localization of these states, and their dispersions, was not investigated theoretically before, and our results contribute to the general understanding of the resonant states in semiconductors. 

In addition, as introduction of the transition metal to the semiconductor  {may strongly modify its magnetic properties (e.g. Cr doped PbTe\cite{Grodzicka1994,michele-pbtecr,story-pbtecr}), sometimes even resulting in forming an ordered ferromagnetic state (so called diluted magnetic semiconductors, DMS\cite{dms-dietl-rmp,dms-theory-rmp}), the magnetic ground state is studied as well, and we find that local magnetic moment appears on Ti atoms, both in ordered (ferromagnetic) as well as disordered (paramagnetic-like) phases.}

{\it Results and discussion}
Electronic structure calculations for Ti doped PbTe were performed using the Korringa-Kohn-Rostoker method, and applying the coherent potential approximation (CPA) to account for the chemical disorder. The Munich SPRKKR\cite{sprkkr,ebert-kkr2011} package was used, the details of the calculations were similar to those presented in Ref.~\cite{bw2013}. The experimental crystal structure\cite{ravich-book} (NaCl type, space group no 225) was used, with the lattice parameter of 6.46 \AA, and with Ti placed on Pb sites. Calculations in the relativistic (including spin-orbit interaction), as well as scalar-relativistic modes were done, relativistic results are presented here. 

\begin{figure}[b]
\includegraphics[width=0.35\textwidth]{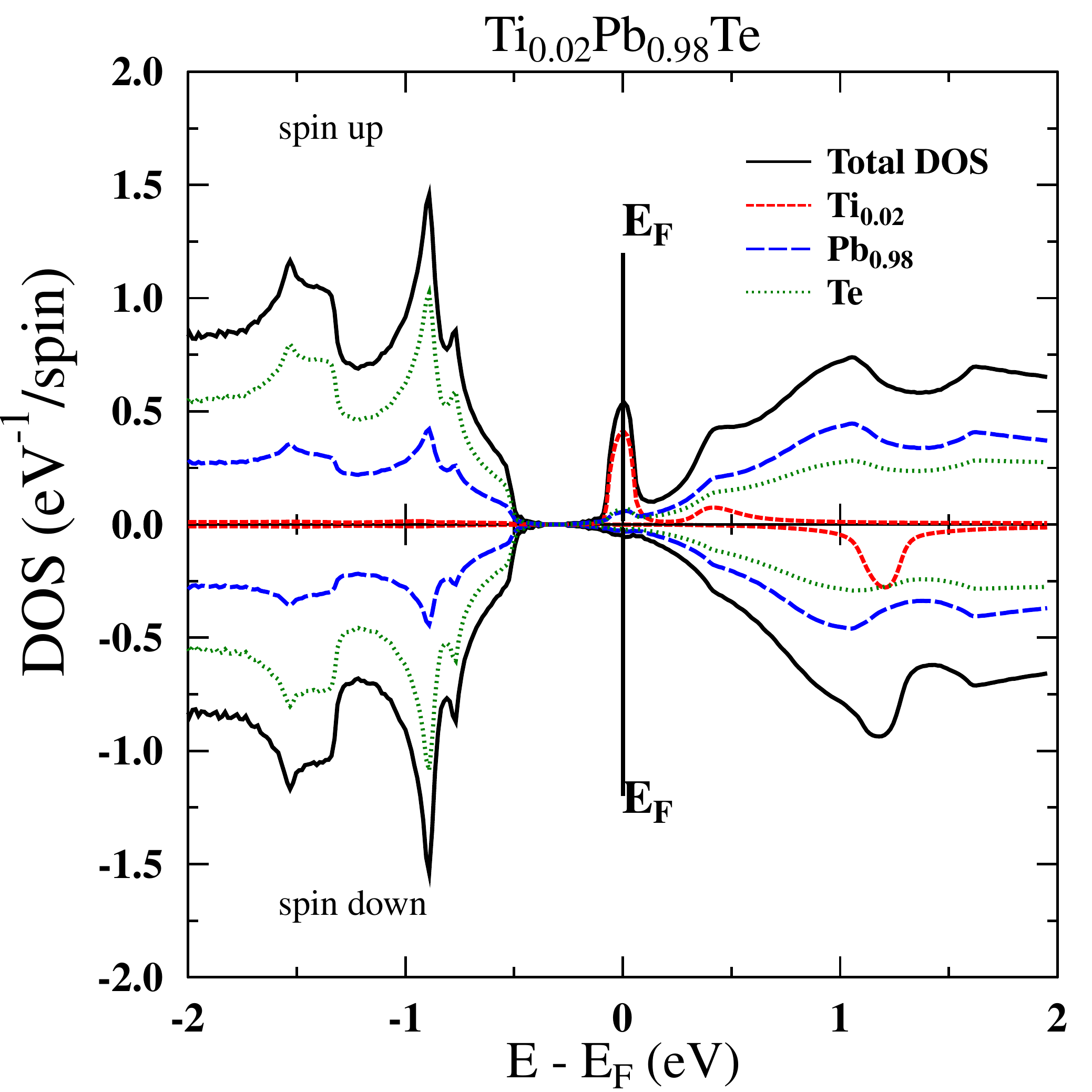}
\caption{KKR-CPA density of states for Ti$_{0.02}$Pb$_{0.98}$Te. Black solid line is the total density of states, color lines show the contribution to DOS from constituent atoms.}
\label{fig:1}
\end{figure}
 
\begin{figure}[t]
\includegraphics[width=0.35\textwidth]{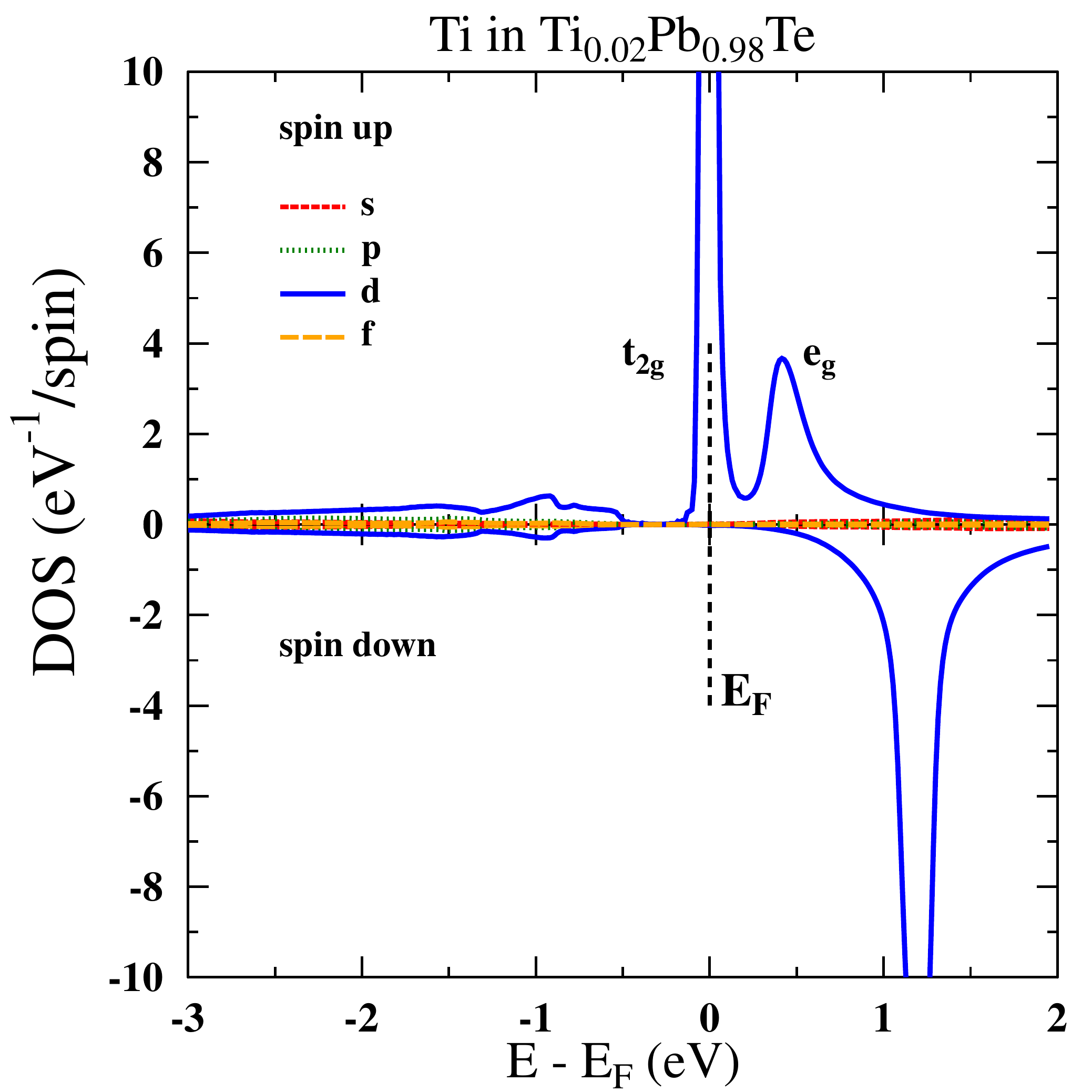}
\caption{KKR-CPA density of states of Ti atom in Ti$_{0.02}$Pb$_{0.98}$Te. Contributions from s, p, d, and f states are marked by colors.}
\label{fig:2}
\end{figure}

Electronic densities of states for Ti$_{0.02}$Pb$_{0.98}$Te are presented in Fig.~\ref{fig:1}, whereas the site-decomposed DOS for Ti atom, with angular momentum decomposition, is shown in Fig.~\ref{fig:2}. First of all, the calculations predict the location of the Fermi level in the conduction band of PbTe, in agreement with the experimental n-type conductivity of the system.\cite{konig-pbte_ti} The integrated density of states from the conduction band edge up to the Fermi level shows that there are about 0.04 electrons per formula unit (about 0.035 up and 0.005 down).
For the 2\% concentration of Ti atoms, this means that Titanium acts as a two electron donor, which would correspond to Ti$^{+4}$ atomic configuration. However, the occupied electronic states between conduction band edge and $E_F$ do not exclusively come from two 3d electrons of Ti, since part of the 3d Ti states are located in the valence band below the gap. The resonant-like character of Ti atom in PbTe is seen, as it creates sharp peak at the Fermi level.
The DOS shows spin polarization, predicting the formation of magnetic ground state in the system. Magnetic moment per formula unit, in agreement with the number of spin-up and spin-down states located above the band gap, is about 0.03 $\mu_B$. Magnetic moments appear on Titanium atoms, which is not very common case (but Ti was found to be magnetic impurity in other systems, e.g. in semiconducting ZnO).\cite{fm_tizno-prl} The magnetic moment comes mainly from the spin-polarized 3d states, which forms three peaks in the conduction band DOS, as can be seen in Fig.~\ref{fig:2}. Spin-up DOS shows clear t$_{2g}$ -- e$_{g}$ splitting, whereas for spin-down states, t$_{2g}$ and e$_{g}$ (empty) states overlap. Fermi level is pinned in the middle of the first, spin-up DOS peak (t$_{2g}$), hence there should be around 1.5 spin-up 3d Ti electrons in the conduction band (actual number is 1.3 from the DOS integration). The calculated total spin magnetic moment per Ti atom is $\mu_{Ti} = 1.65 \mu_B$, with the 0.35~$\mu_B$ 
contribution from the 3d 
states located in the valence band. This small positive magnetization of Ti in the valence band (equal to 0.007 $\mu_B$ per f.u.) is compensated by the small negative moments from valence states of Pb and Te atoms, giving the zero contribution to magnetization from the filled valence band, as expected. On the other hand, in the conduction band, Pb and Te states exhibit small positive magnetization, supplying the 'missing' 0.004~$\mu_B$ to the total magnetization of the conduction band and hence the magnetic moment per formula unit ($0.02\times 1.3$~$\mu_B$ from Ti + $0.004$~$\mu_B$ from Pb and Te = $0.03$~$\mu_B$ per f.u., in agreement with the number derived from the integration of the total DOS).
In addition to the spin moment, orbital magnetic moment on Ti equal to $\mu_{orb} = -0.24$~$\mu_{B}$ was found. 

\begin{figure}
\includegraphics[width=0.48\textwidth]{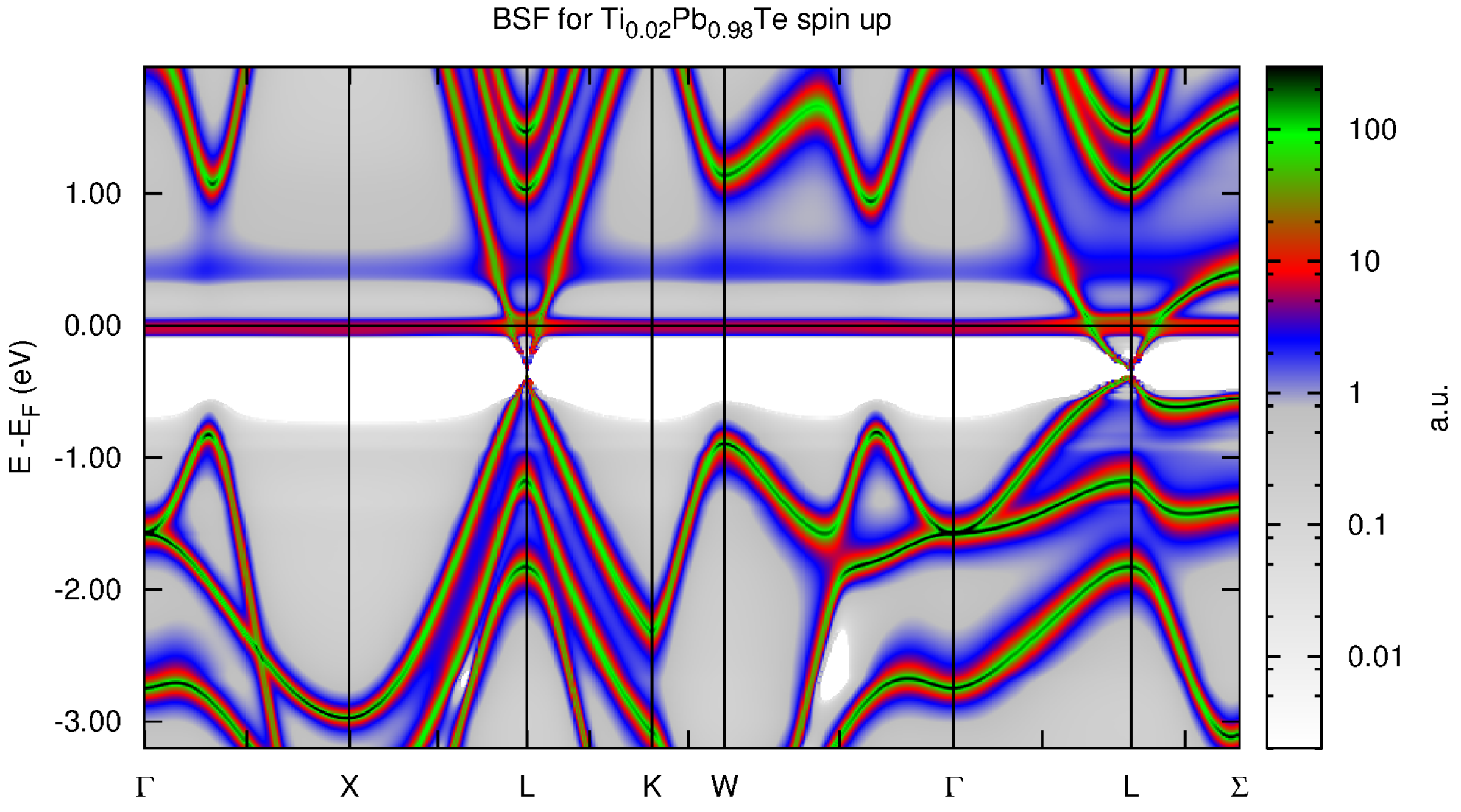}\\
\includegraphics[width=0.48\textwidth]{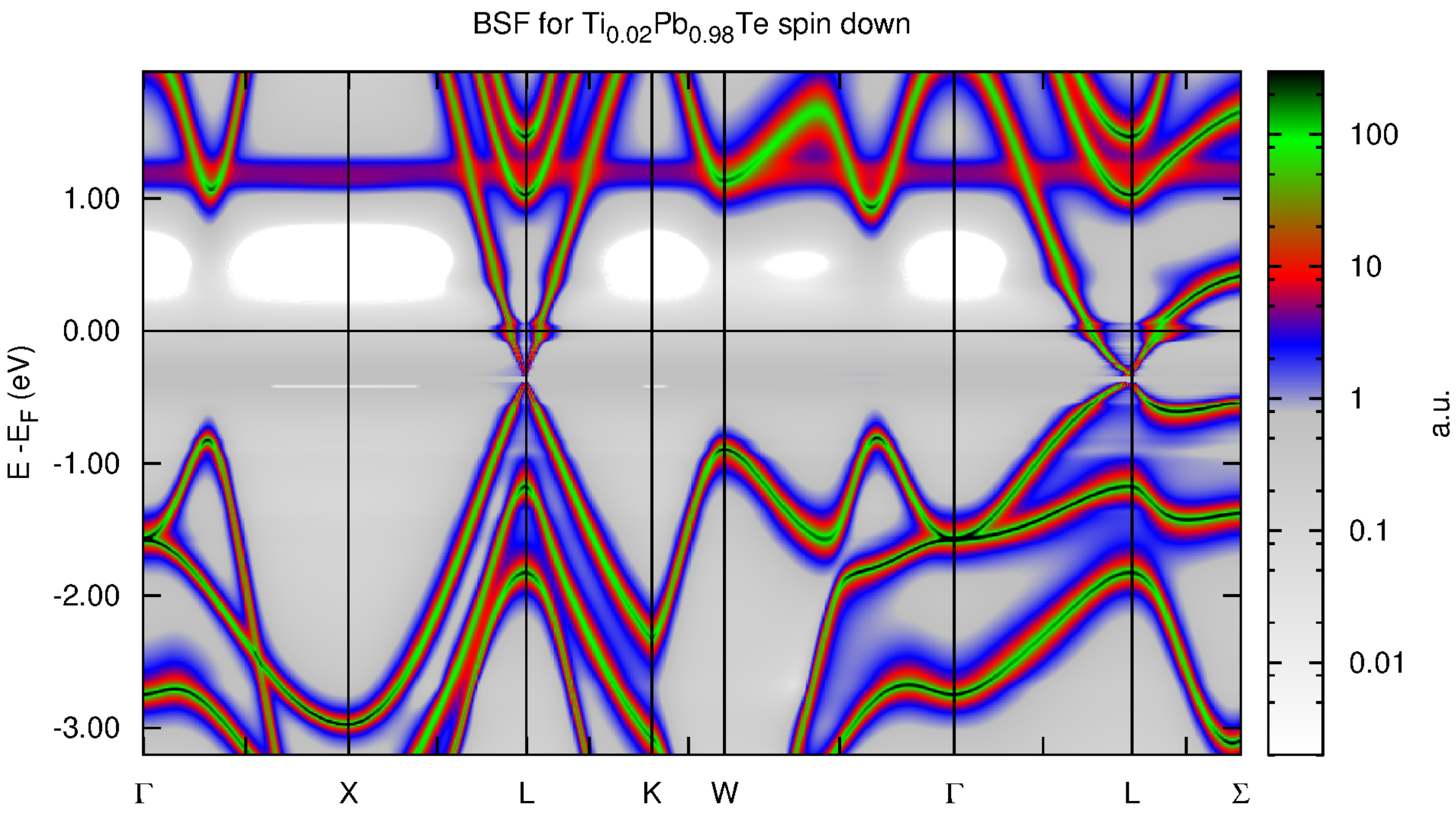}
\caption{The two-dimensional projections of Bloch spectral functions of Ti$_{0.02}$Pb$_{0.98}$Te in high-symmetry directions, for spin up and spin down states. Black color corresponds to BSF values greater than 300 a.u.}
\label{fig:3}
\end{figure}

To verify whether the appearance and magnitude of the magnetic moment on Ti atom might be related to the assumed type of magnetic ordering (which is naturally ferro for a single unit cell with periodic boundary conditions), calculations for the disordered paramagnetic-like system were done. To simulate the complete magnetic disorder, so called disordered local moments (DLM) method\cite{dlm1,dlm2,bw-fe2p} was used. In these calculations, the titanium atoms with initial 'up' and 'down' magnetic moments were placed on the Pb site, simulating Pb$_{0.98}$Ti$_{0.01}^\uparrow$Ti$_{0.01}^\downarrow$Te system with the use of the coherent potential approximation. 
After reaching convergence in the self-consistent cycle, the magnetic moment on Ti atoms were equal to $\pm 1.68$~$mu_B$, thus remained practically unchanged from the starting values, and canceling each other to the zero total magnetic moment of the unit cell. Thus we may expect, that even if the magnetic interactions between Ti ions in the real system are not strong enough to develop the magnetically ordered state, Ti atoms may posses magnetic moments leading to semimagnetic state. 

 {The magnetic properties of PbTe:Ti were unfortunately not studied experimentally, so direct comparison with experiment is not possible at the moment, and the question whether in PbTe:Ti any ordered or disordered magnetic state is formed, remains open. Since the hybridization between Ti $d$ states and $p$ states of PbTe is rather small (see, below) the Zener's $p-d$ exchange mechanism, responsible for ferromagnetism e.g. in Ga$_{1-x}$Mn$_x$As\cite{dietl-science} is not expected, however other mechanisms, like double exchange\cite{dms-theory-rmp} cannot be excluded. }

Now let us discuss the modifications of the electronic bands to address the problem of localization of the impurity states. To show how presence of Ti modifies the electronic dispersion relations of PbTe, Bloch spectral density functions $A^B({\bf k}, E)$ (BSF) were calculated. To recall shortly, BSF can be considered as a generalization of the dispersion relations for the disordered solid.\cite{ebert-kkr2011,faulkner-cpa,Ebert-bloch} 
For the ordered crystal without impurities, $A^B({\bf k}, E)$ at given ${\bf k}$ is a Dirac delta function of energy, being non-zero only in $(k,E)$ points, where electron in the band $\nu$ has an energy eigenvalue $E_{\nu,\mathbf{k}}$, thus $A^B(\mathbf{k},E) = \sum_{\nu} \delta(E - E_{\nu,\mathbf{k}})$. In such a case, the series of BSF describe the usual dispersion relations $E(\mathbf{k})$.
In disordered systems, like crystal with substituted atoms, due to the disorder-induced alloy scattering of electrons the electronic bands are smeared, and, for majority of cases, BSF at given ${\bf k}$ takes the form of the Lorentz function, where the peak of BSF gives the position to the center of the band $E_{\nu,\mathbf{k}}$, and full width at half maximum (FWHM) value $\Gamma$ corresponds to the life time of the electronic state\cite{gyorffy-cu_ni}, $\tau = \hbar/\Gamma$.

\begin{figure}
\includegraphics[width=0.48\textwidth]{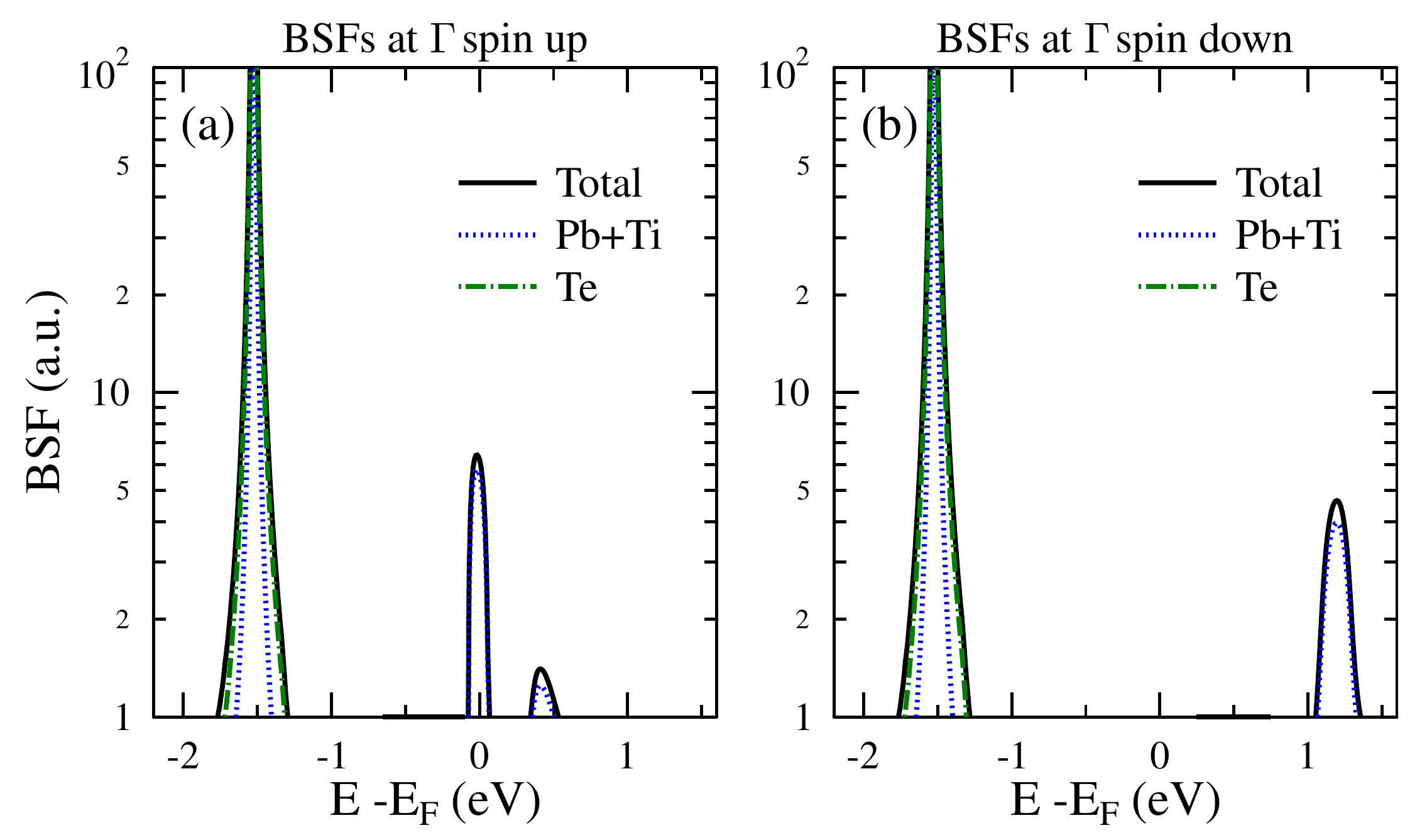}
\caption{Bloch spectral functions at $\Gamma$ point. Color lines show contributions from crystal sites: (4a) occupied by Pb and Ti, and (4b) occupied by Te. The two peaks above 0 eV for spin-up and one for spin-down channels originate from the impurity states, and show no hybridization with Te states. The peak around -1.5 eV comes from the original PbTe band.}
\label{fig:4}
\end{figure}

To visualize the dispersion relations using BSF, two dimensional projections of the BSF for Ti$_{0.02}$Pb$_{0.98}$Te are presented in Fig.~\ref{fig:3}. The color marks the value of BSF, black corresponds to values greater than 300 a.u. The most important feature of presented spectral functions is the dispersionless, flat part of the spin-up BSF, located at the Fermi level. This constant and k-independent part of BSF resembles the flat impurity band, showing that those states are likely to be localized in real space. 
This flat BSF part corresponds to the first spin-up DOS peak in Ti states, presented in the Fig.~\ref{fig:2}. The part of BSF corresponding to the second spin-up DOS peak, in agreement with its larger width and smaller top value, shows smaller intensity. On the spin-down BSF plot, we also clearly recognize similar, flat BSF part, below -1~eV, corresponding to the spin-down DOS peak seen in the Fig.~\ref{fig:2}. To confirm the impurity character of these localized states, and small hybridization of 
the impurity states with the neighboring Te atoms, BSF at $\bf k = (0,0,0)$ ($\Gamma$ point) for spin up and spin down states are plotted in the Fig.~\ref{fig:4}, with spatial decomposition from two crystal sites of the PbTe rock salt structure, i.e. (4a) site, occupied by Pb+Ti, and (4b), occupied by Te. In contrast to the original PbTe band centered about -1.5 eV below $E_F$, showing both sites contribution, the impurity-induced states appearing at 0 eV, 0.5 eV (spin-up), and 1.2 eV (spin-down) show no Te contribution, i.e. no hybridization with nearest neighbors atoms. Since at $\Gamma$ point at this energy in PbTe there are no electronic bands (cf. Fig.~\ref{fig:3}), the origin of this states is exclusively the presence of Ti impurities.

This localized behavior of Ti impurity remains in agreement with the experimental observations, where Ti was found to create a resonant 
level in the conduction band of PbTe\cite{konig-pbte_ti}, but it did not contribute to the carrier concentration, measured using Hall effect, and did not increase the thermopower at given carrier concentration, comparing to other classical n-type impurities.  {No possitive effect on TE performance due to modifications in electron scattering was also observed.\cite{konig-pbte_ti}}

The theoretical results obtained here for PbTe:Ti case can be now compared to the PbTe:Tl case, since Tl is a resonant impurity in PbTe as well, where increase in thermopower against other impurities, was observed.\cite{heremans-science,ees-review,ees-jaworski,pbte-tl-si-nano} Basing on similar analysis of the Bloch spectral functions, it was recently showed\cite{bw2013} that Thallium states in PbTe did not form any narrow-impurity-band-like features in BSF. Presence of resonant Tl impurity lead to substantial blurring of the highest valence band, mainly between $\Sigma$ and $L$ points in the Brillouin zone (BZ), and the strongly hybridized impurity and host (mainly Te) electronic states formed a 'cloud' of additional electronic states around this band, which was the origin of the peak in density of states. In view of those results, the effect of increase in thermopower by the presence of resonant impurity was analyzed as similar to the effect of band convergence or increase in the band degeneracy, known to 
improve TE properties of materials. 
Now, contrast between the Bloch spectral functions in Ti and Tl doped PbTe gives further support for this reasoning. Creation of the isolated (weekly hybridized) impurity-band-like states (Ti case), instead of adding of extra hybridized states around existing band (Tl case) results in no increase in Seebeck coefficient. 
As was explained e.g. in Refs~\cite{ees-review,bw2013}, even, if the impurity band, having large effective mass, has a large partial thermopower $S_{imp}$, due to its localized character will have very small, partial conductivity $\sigma_{imp}$, much smaller than the host material conductivity $\sigma_{host}$. Thus, for such a two band system, where the total thermopower is the weighted average $S \simeq (\sigma_{host}S_{host} + \sigma_{imp}S_{imp})/(\sigma_{host}+\sigma_{imp})$, condition $\sigma_{host} \gg \sigma_{imp}$ results in $S \simeq S_{host}$.
This once again confirms that the delocalization and hybridization of the resonant impurity states is the key parameter to improve thermoelectric properties of semiconductors.

{\it Summary}
Results of the first principles calculations of electronic structure of Ti$_{0.02}$Pb$_{0.98}$Te were presented. Analysis of densities of states showed that Ti is a donor in PbTe. Formation of local magnetic moments equal to about 1.6 $\mu_B$ per Ti atom was found, even in the absence of magnetic order.
Calculated Bloch spectral functions showed, that 3d electrons of Ti form dispersionless impurity-band-like states in the conduction band of PbTe, poorly hybridized with the host crystal's electronic states. These observations were interpreted as a signature of localization of Ti states, in agreement with experimental studies. The comparison of the BSF obtained here for Ti$_{0.02}$Pb$_{0.98}$Te with those published for Tl$_{0.02}$Pb$_{0.98}$Te give intuitive explanation for the different effects of Tl and Ti resonant impurities on thermopower of PbTe, i.e. increase in first case and no effect in second case.

{\it Acknowledgments}
The work at the AGH-UST, Krakow, Poland, was supported by the Polish National Science Center (project no. DEC-2011/02/A/ST3/00124). The work at the OSU, Columbus, Ohio, was supported by AFOSR MURI ''Cryogenic Peltier Cooling'', contract FA9550-10-1-0533.


\begin{thebibliography}{29}
\expandafter\ifx\csname natexlab\endcsname\relax\def\natexlab#1{#1}\fi
\expandafter\ifx\csname bibnamefont\endcsname\relax
  \def\bibnamefont#1{#1}\fi
\expandafter\ifx\csname bibfnamefont\endcsname\relax
  \def\bibfnamefont#1{#1}\fi
\expandafter\ifx\csname citenamefont\endcsname\relax
  \def\citenamefont#1{#1}\fi
\expandafter\ifx\csname url\endcsname\relax
  \def\url#1{\texttt{#1}}\fi
\expandafter\ifx\csname urlprefix\endcsname\relax\def\urlprefix{URL }\fi
\providecommand{\bibinfo}[2]{#2}
\providecommand{\eprint}[2][]{\url{#2}}

\bibitem[{\citenamefont{Heremans et~al.}(2012)\citenamefont{Heremans,
  Wiendlocha, and Chamoire}}]{ees-review}
\bibinfo{author}{\bibfnamefont{J.~P.} \bibnamefont{Heremans}},
  \bibinfo{author}{\bibfnamefont{B.}~\bibnamefont{Wiendlocha}},
  \bibnamefont{and} \bibinfo{author}{\bibfnamefont{A.~M.}
  \bibnamefont{Chamoire}}, \bibinfo{journal}{Energy Environ. Sci.}
  \textbf{\bibinfo{volume}{5}}, \bibinfo{pages}{5510} (\bibinfo{year}{2012}),
  \urlprefix\url{http://dx.doi.org/10.1039/C1EE02612G}.

\bibitem[{\citenamefont{Wiendlocha}(2013)}]{bw2013}
\bibinfo{author}{\bibfnamefont{B.}~\bibnamefont{Wiendlocha}},
  \bibinfo{journal}{Phys. Rev. B} \textbf{\bibinfo{volume}{88}},
  \bibinfo{pages}{205205} (\bibinfo{year}{2013}),
  \urlprefix\url{http://link.aps.org/doi/10.1103/PhysRevB.88.205205}.

\bibitem[{\citenamefont{Nemov and {Yu. I. Ravich}}(1998)}]{ravich-review}
\bibinfo{author}{\bibfnamefont{S.~A.} \bibnamefont{Nemov}} \bibnamefont{and}
  \bibinfo{author}{\bibnamefont{{Yu. I. Ravich}}}, \bibinfo{journal}{Physics -
  Uspekhi} \textbf{\bibinfo{volume}{41}}, \bibinfo{pages}{735}
  (\bibinfo{year}{1998}).

\bibitem[{\citenamefont{Kaidanov and {Yu. I. Ravich}}(1985)}]{ravich-review2}
\bibinfo{author}{\bibfnamefont{V.~I.} \bibnamefont{Kaidanov}} \bibnamefont{and}
  \bibinfo{author}{\bibnamefont{{Yu. I. Ravich}}}, \bibinfo{journal}{Sov. Phys.
  Usp.} \textbf{\bibinfo{volume}{28}}, \bibinfo{pages}{31}
  (\bibinfo{year}{1985}).

\bibitem[{\citenamefont{Heremans et~al.}(2008)\citenamefont{Heremans, Jovovic,
  Toberer, Saramat, Kurosaki, Charoenphakdee, Yamanaka, and
  Snyder}}]{heremans-science}
\bibinfo{author}{\bibfnamefont{J.~P.} \bibnamefont{Heremans}},
  \bibinfo{author}{\bibfnamefont{V.}~\bibnamefont{Jovovic}},
  \bibinfo{author}{\bibfnamefont{E.~S.} \bibnamefont{Toberer}},
  \bibinfo{author}{\bibfnamefont{A.}~\bibnamefont{Saramat}},
  \bibinfo{author}{\bibfnamefont{K.}~\bibnamefont{Kurosaki}},
  \bibinfo{author}{\bibfnamefont{A.}~\bibnamefont{Charoenphakdee}},
  \bibinfo{author}{\bibfnamefont{S.}~\bibnamefont{Yamanaka}}, \bibnamefont{and}
  \bibinfo{author}{\bibfnamefont{G.~J.} \bibnamefont{Snyder}},
  \bibinfo{journal}{Science} \textbf{\bibinfo{volume}{321}},
  \bibinfo{pages}{554} (\bibinfo{year}{2008}),
  \eprint{http://www.sciencemag.org/content/321/5888/554.full.pdf},
  \urlprefix\url{http://www.sciencemag.org/content/321/5888/554.abstract}.

\bibitem[{\citenamefont{Jaworski et~al.}(2011)\citenamefont{Jaworski,
  Wiendlocha, Jovovic, and Heremans}}]{ees-jaworski}
\bibinfo{author}{\bibfnamefont{C.~M.} \bibnamefont{Jaworski}},
  \bibinfo{author}{\bibfnamefont{B.}~\bibnamefont{Wiendlocha}},
  \bibinfo{author}{\bibfnamefont{V.}~\bibnamefont{Jovovic}}, \bibnamefont{and}
  \bibinfo{author}{\bibfnamefont{J.~P.} \bibnamefont{Heremans}},
  \bibinfo{journal}{Energy Environ. Sci.} \textbf{\bibinfo{volume}{4}},
  \bibinfo{pages}{4155} (\bibinfo{year}{2011}),
  \urlprefix\url{http://dx.doi.org/10.1039/C1EE01895G}.

\bibitem[{\citenamefont{Zhang et~al.}(2012)\citenamefont{Zhang, Wang, Zhang,
  Liu, Yu, Wang, Wang, Ni, Chen, and Ren}}]{pbte-tl-si-nano}
\bibinfo{author}{\bibfnamefont{Q.}~\bibnamefont{Zhang}},
  \bibinfo{author}{\bibfnamefont{H.}~\bibnamefont{Wang}},
  \bibinfo{author}{\bibfnamefont{Q.}~\bibnamefont{Zhang}},
  \bibinfo{author}{\bibfnamefont{W.}~\bibnamefont{Liu}},
  \bibinfo{author}{\bibfnamefont{B.}~\bibnamefont{Yu}},
  \bibinfo{author}{\bibfnamefont{H.}~\bibnamefont{Wang}},
  \bibinfo{author}{\bibfnamefont{D.}~\bibnamefont{Wang}},
  \bibinfo{author}{\bibfnamefont{G.}~\bibnamefont{Ni}},
  \bibinfo{author}{\bibfnamefont{G.}~\bibnamefont{Chen}}, \bibnamefont{and}
  \bibinfo{author}{\bibfnamefont{Z.}~\bibnamefont{Ren}}, \bibinfo{journal}{Nano
  Letters} \textbf{\bibinfo{volume}{12}}, \bibinfo{pages}{2324}
  (\bibinfo{year}{2012}),
  \eprint{http://pubs.acs.org/doi/pdf/10.1021/nl3002183},
  \urlprefix\url{http://pubs.acs.org/doi/abs/10.1021/nl3002183}.

\bibitem[{\citenamefont{K\"onig et~al.}(2011)\citenamefont{K\"onig, Nielsen,
  Gao, Winkler, Jacquot, B\"ottner, and Heremans}}]{konig-pbte_ti}
\bibinfo{author}{\bibfnamefont{J.~D.} \bibnamefont{K\"onig}},
  \bibinfo{author}{\bibfnamefont{M.~D.} \bibnamefont{Nielsen}},
  \bibinfo{author}{\bibfnamefont{Y.-B.} \bibnamefont{Gao}},
  \bibinfo{author}{\bibfnamefont{M.}~\bibnamefont{Winkler}},
  \bibinfo{author}{\bibfnamefont{A.}~\bibnamefont{Jacquot}},
  \bibinfo{author}{\bibfnamefont{H.}~\bibnamefont{B\"ottner}},
  \bibnamefont{and} \bibinfo{author}{\bibfnamefont{J.~P.}
  \bibnamefont{Heremans}}, \bibinfo{journal}{Phys. Rev. B}
  \textbf{\bibinfo{volume}{84}}, \bibinfo{pages}{205126}
  (\bibinfo{year}{2011}),
  \urlprefix\url{http://link.aps.org/doi/10.1103/PhysRevB.84.205126}.

\bibitem[{\citenamefont{Paul and Banerji}(2011)}]{pbte_cr_jap}
\bibinfo{author}{\bibfnamefont{B.}~\bibnamefont{Paul}} \bibnamefont{and}
  \bibinfo{author}{\bibfnamefont{P.}~\bibnamefont{Banerji}},
  \bibinfo{journal}{Journal of Applied Physics} \textbf{\bibinfo{volume}{109}},
  \bibinfo{eid}{103710} (\bibinfo{year}{2011}),
  \urlprefix\url{http://scitation.aip.org/content/aip/journal/jap/109/10/10.10%
63/1.3592349}.

\bibitem[{\citenamefont{Paul et~al.}(2011)\citenamefont{Paul, Rawat, and
  Banerji}}]{pbte_cr_apl}
\bibinfo{author}{\bibfnamefont{B.}~\bibnamefont{Paul}},
  \bibinfo{author}{\bibfnamefont{P.~K.} \bibnamefont{Rawat}}, \bibnamefont{and}
  \bibinfo{author}{\bibfnamefont{P.}~\bibnamefont{Banerji}},
  \bibinfo{journal}{Applied Physics Letters} \textbf{\bibinfo{volume}{98}},
  \bibinfo{eid}{262101} (\bibinfo{year}{2011}),
  \urlprefix\url{http://scitation.aip.org/content/aip/journal/apl/98/26/10.106%
3/1.3603962}.

\bibitem[{\citenamefont{Grodzicka et~al.}(1994)\citenamefont{Grodzicka,
  Dobrowolski, Kossut, Story, and Witkowska}}]{Grodzicka1994}
\bibinfo{author}{\bibfnamefont{E.}~\bibnamefont{Grodzicka}},
  \bibinfo{author}{\bibfnamefont{W.}~\bibnamefont{Dobrowolski}},
  \bibinfo{author}{\bibfnamefont{J.}~\bibnamefont{Kossut}},
  \bibinfo{author}{\bibfnamefont{T.}~\bibnamefont{Story}}, \bibnamefont{and}
  \bibinfo{author}{\bibfnamefont{B.}~\bibnamefont{Witkowska}},
  \bibinfo{journal}{Journal of Crystal Growth} \textbf{\bibinfo{volume}{138}},
  \bibinfo{pages}{1034 } (\bibinfo{year}{1994}), ISSN
  \bibinfo{issn}{0022-0248},
  \urlprefix\url{http://www.sciencedirect.com/science/article/pii/002202489490%
9512}.

\bibitem[{\citenamefont{Nielsen et~al.}(2012)\citenamefont{Nielsen, Levin,
  Jaworski, Schmidt-Rohr, and Heremans}}]{michele-pbtecr}
\bibinfo{author}{\bibfnamefont{M.~D.} \bibnamefont{Nielsen}},
  \bibinfo{author}{\bibfnamefont{E.~M.} \bibnamefont{Levin}},
  \bibinfo{author}{\bibfnamefont{C.~M.} \bibnamefont{Jaworski}},
  \bibinfo{author}{\bibfnamefont{K.}~\bibnamefont{Schmidt-Rohr}},
  \bibnamefont{and} \bibinfo{author}{\bibfnamefont{J.~P.}
  \bibnamefont{Heremans}}, \bibinfo{journal}{Phys. Rev. B}
  \textbf{\bibinfo{volume}{85}}, \bibinfo{pages}{045210}
  (\bibinfo{year}{2012}),
  \urlprefix\url{http://link.aps.org/doi/10.1103/PhysRevB.85.045210}.

\bibitem[{\citenamefont{Story et~al.}(1992)\citenamefont{Story, Grodzicka,
  Witkowska, G\'orecka, and Dobrowolski}}]{story-pbtecr}
\bibinfo{author}{\bibfnamefont{T.}~\bibnamefont{Story}},
  \bibinfo{author}{\bibfnamefont{E.}~\bibnamefont{Grodzicka}},
  \bibinfo{author}{\bibfnamefont{B.}~\bibnamefont{Witkowska}},
  \bibinfo{author}{\bibfnamefont{J.}~\bibnamefont{G\'orecka}},
  \bibnamefont{and}
  \bibinfo{author}{\bibfnamefont{W.}~\bibnamefont{Dobrowolski}},
  \bibinfo{journal}{Acta Phys. Pol. A} \textbf{\bibinfo{volume}{82}},
  \bibinfo{pages}{879} (\bibinfo{year}{1992}),
  \urlprefix\url{http://przyrbwn.icm.edu.pl/APP/ABSTR/82/a82-5-40.html}.

\bibitem[{\citenamefont{Morelli et~al.}(2003)\citenamefont{Morelli, Heremans,
  and Thrush}}]{pbte_fe-heremans}
\bibinfo{author}{\bibfnamefont{D.~T.} \bibnamefont{Morelli}},
  \bibinfo{author}{\bibfnamefont{J.~P.} \bibnamefont{Heremans}},
  \bibnamefont{and} \bibinfo{author}{\bibfnamefont{C.~M.}
  \bibnamefont{Thrush}}, \bibinfo{journal}{Phys. Rev. B}
  \textbf{\bibinfo{volume}{67}}, \bibinfo{pages}{035206}
  (\bibinfo{year}{2003}),
  \urlprefix\url{http://link.aps.org/doi/10.1103/PhysRevB.67.035206}.

\bibitem[{\citenamefont{Skipetrov
  et~al.}(2014{\natexlab{a}})\citenamefont{Skipetrov, Kruleveckaya, Skipetrova,
  Slynko, and Slynko}}]{pbte_fe_press}
\bibinfo{author}{\bibfnamefont{E.~P.} \bibnamefont{Skipetrov}},
  \bibinfo{author}{\bibfnamefont{O.~V.} \bibnamefont{Kruleveckaya}},
  \bibinfo{author}{\bibfnamefont{L.~A.} \bibnamefont{Skipetrova}},
  \bibinfo{author}{\bibfnamefont{E.~I.} \bibnamefont{Slynko}},
  \bibnamefont{and} \bibinfo{author}{\bibfnamefont{V.~E.}
  \bibnamefont{Slynko}}, \bibinfo{journal}{Applied Physics Letters}
  \textbf{\bibinfo{volume}{105}}, \bibinfo{eid}{022101}
  (\bibinfo{year}{2014}{\natexlab{a}}),
  \urlprefix\url{http://scitation.aip.org/content/aip/journal/apl/105/2/10.106%
3/1.4890381}.

\bibitem[{\citenamefont{Skipetrov
  et~al.}(2014{\natexlab{b}})\citenamefont{Skipetrov, Skipetrova, Knotko,
  Slynko, and Slynko}}]{pbte_sc}
\bibinfo{author}{\bibfnamefont{E.~P.} \bibnamefont{Skipetrov}},
  \bibinfo{author}{\bibfnamefont{L.~A.} \bibnamefont{Skipetrova}},
  \bibinfo{author}{\bibfnamefont{A.~V.} \bibnamefont{Knotko}},
  \bibinfo{author}{\bibfnamefont{E.~I.} \bibnamefont{Slynko}},
  \bibnamefont{and} \bibinfo{author}{\bibfnamefont{V.~E.}
  \bibnamefont{Slynko}}, \bibinfo{journal}{Journal of Applied Physics}
  \textbf{\bibinfo{volume}{115}}, \bibinfo{eid}{133702}
  (\bibinfo{year}{2014}{\natexlab{b}}),
  \urlprefix\url{http://scitation.aip.org/content/aip/journal/jap/115/13/10.10%
63/1.4870230}.

\bibitem[{\citenamefont{Dietl and Ohno}(2014)}]{dms-dietl-rmp}
\bibinfo{author}{\bibfnamefont{T.}~\bibnamefont{Dietl}} \bibnamefont{and}
  \bibinfo{author}{\bibfnamefont{H.}~\bibnamefont{Ohno}},
  \bibinfo{journal}{Rev. Mod. Phys.} \textbf{\bibinfo{volume}{86}},
  \bibinfo{pages}{187} (\bibinfo{year}{2014}),
  \urlprefix\url{http://link.aps.org/doi/10.1103/RevModPhys.86.187}.

\bibitem[{\citenamefont{Sato et~al.}(2010)\citenamefont{Sato, Bergqvist,
  Kudrnovsk\'y, Dederichs, Eriksson, Turek, Sanyal, Bouzerar, Katayama-Yoshida,
  Dinh et~al.}}]{dms-theory-rmp}
\bibinfo{author}{\bibfnamefont{K.}~\bibnamefont{Sato}},
  \bibinfo{author}{\bibfnamefont{L.}~\bibnamefont{Bergqvist}},
  \bibinfo{author}{\bibfnamefont{J.}~\bibnamefont{Kudrnovsk\'y}},
  \bibinfo{author}{\bibfnamefont{P.~H.} \bibnamefont{Dederichs}},
  \bibinfo{author}{\bibfnamefont{O.}~\bibnamefont{Eriksson}},
  \bibinfo{author}{\bibfnamefont{I.}~\bibnamefont{Turek}},
  \bibinfo{author}{\bibfnamefont{B.}~\bibnamefont{Sanyal}},
  \bibinfo{author}{\bibfnamefont{G.}~\bibnamefont{Bouzerar}},
  \bibinfo{author}{\bibfnamefont{H.}~\bibnamefont{Katayama-Yoshida}},
  \bibinfo{author}{\bibfnamefont{V.~A.} \bibnamefont{Dinh}},
  \bibnamefont{et~al.}, \bibinfo{journal}{Rev. Mod. Phys.}
  \textbf{\bibinfo{volume}{82}}, \bibinfo{pages}{1633} (\bibinfo{year}{2010}),
  \urlprefix\url{http://link.aps.org/doi/10.1103/RevModPhys.82.1633}.

\bibitem[{\citenamefont{{Ebert et al.}}(2012)}]{sprkkr}
\bibinfo{author}{\bibfnamefont{H.}~\bibnamefont{{Ebert et al.}}},
  \bibinfo{journal}{The Munich SPR-KKR package, version 6.3.1.
  http://ebert.cup.uni-muenchen.de/SPRKKR}  (\bibinfo{year}{2012}).

\bibitem[{\citenamefont{{Ebert} et~al.}(2011)\citenamefont{{Ebert},
  {K{\"o}dderitzsch}, and {Min{\'a}r}}}]{ebert-kkr2011}
\bibinfo{author}{\bibfnamefont{H.}~\bibnamefont{{Ebert}}},
  \bibinfo{author}{\bibfnamefont{D.}~\bibnamefont{{K{\"o}dderitzsch}}},
  \bibnamefont{and}
  \bibinfo{author}{\bibfnamefont{J.}~\bibnamefont{{Min{\'a}r}}},
  \bibinfo{journal}{Reports on Progress in Physics}
  \textbf{\bibinfo{volume}{74}}, \bibinfo{pages}{096501}
  (\bibinfo{year}{2011}).

\bibitem[{\citenamefont{{Yu. I. Ravich} et~al.}(1970)\citenamefont{{Yu. I.
  Ravich}, Efimova, and Smirnov}}]{ravich-book}
\bibinfo{author}{\bibnamefont{{Yu. I. Ravich}}},
  \bibinfo{author}{\bibfnamefont{B.~A.} \bibnamefont{Efimova}},
  \bibnamefont{and} \bibinfo{author}{\bibfnamefont{I.~A.}
  \bibnamefont{Smirnov}}, \emph{\bibinfo{title}{Semiconducting Lead
  Chalcogenides}} (\bibinfo{publisher}{Plenum Press}, \bibinfo{address}{New
  York}, \bibinfo{year}{1970}).

\bibitem[{\citenamefont{Venkatesan et~al.}(2004)\citenamefont{Venkatesan,
  Fitzgerald, Lunney, and Coey}}]{fm_tizno-prl}
\bibinfo{author}{\bibfnamefont{M.}~\bibnamefont{Venkatesan}},
  \bibinfo{author}{\bibfnamefont{C.~B.} \bibnamefont{Fitzgerald}},
  \bibinfo{author}{\bibfnamefont{J.~G.} \bibnamefont{Lunney}},
  \bibnamefont{and} \bibinfo{author}{\bibfnamefont{J.~M.~D.}
  \bibnamefont{Coey}}, \bibinfo{journal}{Phys. Rev. Lett.}
  \textbf{\bibinfo{volume}{93}}, \bibinfo{pages}{177206}
  (\bibinfo{year}{2004}),
  \urlprefix\url{http://link.aps.org/doi/10.1103/PhysRevLett.93.177206}.

\bibitem[{\citenamefont{Pindor et~al.}(1983)\citenamefont{Pindor, Staunton,
  Stocks, and Winter}}]{dlm1}
\bibinfo{author}{\bibfnamefont{A.~J.} \bibnamefont{Pindor}},
  \bibinfo{author}{\bibfnamefont{J.}~\bibnamefont{Staunton}},
  \bibinfo{author}{\bibfnamefont{G.~M.} \bibnamefont{Stocks}},
  \bibnamefont{and} \bibinfo{author}{\bibfnamefont{H.}~\bibnamefont{Winter}},
  \bibinfo{journal}{Journal of Physics F: Metal Physics}
  \textbf{\bibinfo{volume}{13}}, \bibinfo{pages}{979} (\bibinfo{year}{1983}),
  \urlprefix\url{http://stacks.iop.org/0305-4608/13/i=5/a=012}.

\bibitem[{\citenamefont{Gyorffy et~al.}(1985)\citenamefont{Gyorffy, Pindor,
  Staunton, Stocks, and Winter}}]{dlm2}
\bibinfo{author}{\bibfnamefont{B.~L.} \bibnamefont{Gyorffy}},
  \bibinfo{author}{\bibfnamefont{A.~J.} \bibnamefont{Pindor}},
  \bibinfo{author}{\bibfnamefont{J.}~\bibnamefont{Staunton}},
  \bibinfo{author}{\bibfnamefont{G.~M.} \bibnamefont{Stocks}},
  \bibnamefont{and} \bibinfo{author}{\bibfnamefont{H.}~\bibnamefont{Winter}},
  \bibinfo{journal}{Journal of Physics F: Metal Physics}
  \textbf{\bibinfo{volume}{15}}, \bibinfo{pages}{1337} (\bibinfo{year}{1985}),
  \urlprefix\url{http://stacks.iop.org/0305-4608/15/i=6/a=018}.

\bibitem[{\citenamefont{{Wiendlocha} et~al.}(2008)\citenamefont{{Wiendlocha},
  {Tobola}, {Kaprzyk}, {Zach}, {Hlil}, and {Fruchart}}}]{bw-fe2p}
\bibinfo{author}{\bibfnamefont{B.}~\bibnamefont{{Wiendlocha}}},
  \bibinfo{author}{\bibfnamefont{J.}~\bibnamefont{{Tobola}}},
  \bibinfo{author}{\bibfnamefont{S.}~\bibnamefont{{Kaprzyk}}},
  \bibinfo{author}{\bibfnamefont{R.}~\bibnamefont{{Zach}}},
  \bibinfo{author}{\bibfnamefont{E.~K.} \bibnamefont{{Hlil}}},
  \bibnamefont{and}
  \bibinfo{author}{\bibfnamefont{D.}~\bibnamefont{{Fruchart}}},
  \bibinfo{journal}{Journal of Physics D Applied Physics}
  \textbf{\bibinfo{volume}{41}}, \bibinfo{pages}{205007}
  (\bibinfo{year}{2008}), \eprint{0808.0975}.

\bibitem[{\citenamefont{Dietl et~al.}(2000)\citenamefont{Dietl, Ohno,
  Matsukura, Cibert, and Ferrand}}]{dietl-science}
\bibinfo{author}{\bibfnamefont{T.}~\bibnamefont{Dietl}},
  \bibinfo{author}{\bibfnamefont{H.}~\bibnamefont{Ohno}},
  \bibinfo{author}{\bibfnamefont{F.}~\bibnamefont{Matsukura}},
  \bibinfo{author}{\bibfnamefont{J.}~\bibnamefont{Cibert}}, \bibnamefont{and}
  \bibinfo{author}{\bibfnamefont{D.}~\bibnamefont{Ferrand}},
  \bibinfo{journal}{Science} \textbf{\bibinfo{volume}{287}},
  \bibinfo{pages}{1019} (\bibinfo{year}{2000}),
  \eprint{http://www.sciencemag.org/content/287/5455/1019.full.pdf},
  \urlprefix\url{http://www.sciencemag.org/content/287/5455/1019.abstract}.

\bibitem[{\citenamefont{Faulkner and Stocks}(1980)}]{faulkner-cpa}
\bibinfo{author}{\bibfnamefont{J.~S.} \bibnamefont{Faulkner}} \bibnamefont{and}
  \bibinfo{author}{\bibfnamefont{G.~M.} \bibnamefont{Stocks}},
  \bibinfo{journal}{Phys. Rev. B} \textbf{\bibinfo{volume}{21}},
  \bibinfo{pages}{3222} (\bibinfo{year}{1980}),
  \urlprefix\url{http://link.aps.org/doi/10.1103/PhysRevB.21.3222}.

\bibitem[{\citenamefont{Ebert et~al.}(1997)\citenamefont{Ebert, Vernes, and
  Banhart}}]{Ebert-bloch}
\bibinfo{author}{\bibfnamefont{H.}~\bibnamefont{Ebert}},
  \bibinfo{author}{\bibfnamefont{A.}~\bibnamefont{Vernes}}, \bibnamefont{and}
  \bibinfo{author}{\bibfnamefont{J.}~\bibnamefont{Banhart}},
  \bibinfo{journal}{Solid State Communications} \textbf{\bibinfo{volume}{104}},
  \bibinfo{pages}{243} (\bibinfo{year}{1997}), ISSN \bibinfo{issn}{0038-1098},
  \urlprefix\url{http://www.sciencedirect.com/science/article/pii/S00381098970%
02524}.

\bibitem[{\citenamefont{Gordon et~al.}(1981)\citenamefont{Gordon, Temmerman,
  and Gyorffy}}]{gyorffy-cu_ni}
\bibinfo{author}{\bibfnamefont{B.~E.~A.} \bibnamefont{Gordon}},
  \bibinfo{author}{\bibfnamefont{W.~E.} \bibnamefont{Temmerman}},
  \bibnamefont{and} \bibinfo{author}{\bibfnamefont{B.~L.}
  \bibnamefont{Gyorffy}}, \bibinfo{journal}{Journal of Physics F: Metal
  Physics} \textbf{\bibinfo{volume}{11}}, \bibinfo{pages}{821}
  (\bibinfo{year}{1981}),
  \urlprefix\url{http://stacks.iop.org/0305-4608/11/i=4/a=015}.

\end{thebibliography}

\end{document}